\documentclass{article}
\usepackage{graphicx}
\def\ie{i.e.\ }
\def\be{\begin{equation}}
\def\ee{\end{equation}}
\def\ni{\noindent}
\def\pl{\partial}
\def\ep{{\it e-p }}
\def\gb{${\nabla B \; }$}
\def\exb{${E \times B\; }$}
\def\Xint#1{\mathchoice
   {\XXint\displaystyle\textstyle{#1}}%
   {\XXint\textstyle\scriptstyle{#1}}%
   {\XXint\scriptstyle\scriptscriptstyle{#1}}%
   {\XXint\scriptscriptstyle\scriptscriptstyle{#1}}%
   \!\int}
\def\XXint#1#2#3{{\setbox0=\hbox{$#1{#2#3}{\int}$}
     \vcenter{\hbox{$#2#3$}}\kern-.5\wd0}}

\def\dashint{\Xint-}

\begin{document}

\title{Two-Stream Instability Model With Electrons Trapped in Quadrupoles}

\author{Paul J. Channell\\
Los Alamos National Laboratory, Los Alamos, NM 87545 \\
\footnote{This work was supported by the US Department of Energy under Contract Number DE-AC52-06NA25396.}}

\maketitle

\begin{abstract}
We formulate the theory of the two-stream instability (e-cloud instability) with electrons
trapped in quadrupole magnets. We show that a linear instability theory can be sensibly formulated
and analyzed. The growth rates are considerably smaller than the linear growth rates
for the two-stream instability in drift spaces and are close to those actually observed. 
\end{abstract}

\date

\section{Introduction}

The Proton Storage Ring (PSR) at Los Alamos has been troubled for 
some time, \cite{h1}, \cite{h2}, \cite{h3}, \cite{h4}, \cite{h5}, \cite{h6},
 \cite{h7}, \cite{h8}, \cite{h9}, by an instability that
is probably a two-stream instability of the proton beam with background 
electrons, \ie an electron-cloud instability.
We have previously considered the possibility that the instability for a bunched
beam occurred because of electrons in drift regions that were renewed from 
turn to turn, \cite{pjc}; in this case the phase memory of the coherent motion has to reside in
the proton beam which excited the fresh electrons on each turn which then drove
the tail of the proton bunch to larger amplitudes. In this note we will consider instead
the possibility that the instability is
due to electrons that survive from turn to turn. The most likely place in the ring
where the electrons can survive with coherent phase information from turn to turn
is in the quadrupoles, where they are trapped in the magnetic mirrors formed by the
cusp-shaped fields and can drive the \ep instability
in a similar way to free electrons. In this note we will present a simple model of this two-stream instability 
with electrons trapped in quadrupoles.

\subsection{Electron Trapping and Dynamics}

A major assumption of this note is that there are abundant electrons
in the PSR; experimentally this has been observed, though the source
is not completely clear. It is likely that some form of beam induced
multipactor gives rise to the electrons, perhaps initiated by a very
small number of lost beam particles, though other explanations are
possible. 
Normally, one would expect that with a bunched
beam electrons would be expelled during the beam gap and that one could
not have an \ep instability; however, the electrons, however they are
produced, cannot be driven
quickly to the walls in the quadrupoles which act in the transverse
direction as very effective magnetic mirrors. It thus seems possible
that electrons in the quadrupoles could drive the \ep instability. To 
investigate this possibility further
in this section we will make simple estimates of the electron motion in
quadrupoles to establish that electrons can be trapped there for multiple
turns and thus carry coherent phase information to drive the instabillity.
A more accurate investigation of the electron motion in the complex
geometry can and should be done using computer codes, \cite{code}.

The dominant aspect of electron motion in the quads is the rapid rotation
about the magnetic field lines; the cyclotron frequency is

\be
f_c={eB\over 2\pi mc},
\ee

\ni where $e$ is the charge, $B$ is the magnetic field, $m$ is the
mass, and $c$ is the speed of light. For electrons we have

\be
f_c=2.8{\cal B}\,\, {\rm GHz},
\ee

\ni where ${\cal B}$ is the magnetic field measured in kilogauss. Thus, even
very low fields near the axis give rise to cyclotron frequencies that are
hundreds of MHz; most electrons will have cyclotron frequencies that are multiple
GHz. The radius of this rotational motion is, for electrons,

\be
\rho =3.37*10^{-3}{\sqrt {\cal E}\over {\cal B}}\,\, {\rm cm},
\ee

\ni where ${\cal E}$ is the transverse electron energy in eV. Only
very energetic electrons in low field regions will have gyroradii
approaching $1$ cm; most will have gyroradii that are much less than $1$ mm.
Electrons are thus confined transversely to the magnetic field
on cyclotron orbits of small radii and many are confined longitudinally (for electrons)
along the magnetic field
by the increasing magnetic field with radius, \ie by `mirror' confinement.
(Note that longitudinal for the electrons is transverse to the
beam direction.) 

Of course, particles
with large components of velocity parallel to the magnetic field,
\ie those in the `loss cone', are not confined;
presumably these give rise to the electron `tracking' that
has been observed in the quadrupoles.
We will ignore the rapid electron cyclotron motion in the quads and concentrate
on the longitudinal electron mirror motion and transverse
drifts due to electric fields and to
magnetic field non-uniformity, \ie
 a `guiding center' description of the trapped electrons.

In the transverse direction (for electrons) there are three components of electron drift, that due to the
gradient in the magnetic field, the so-called \gb drift,
that due to the field line curvature, 
and that due to any electric fields that are present, the \exb
drift. These drifts give rise to electron velocities
perpendicular to the magnetic field; in fact, in the quads, the
drifts are along the direction of the beam axis and thus can lead to electron loss 
out the ends of the quads.

The \gb drift and curvature velocities are given by

\be
V_{\nabla B}={m(v_{\perp}^2+2 v_{\parallel}^2)\over 2 eB}{\hat b \times \nabla B\over B}c,
\ee

\ni where $\hat b$ is a unit vector in the direction of the magnetic field,
the $v_{\perp}^2$ term is due to the gradient drift and the $v_{\parallel}^2$
term is due to the curvature.
If we assume the parallel and perpendicular electron velocities to be roughly
the same and
adopt the usual model of quadrupole magnetic fields in which a component
is linear in transverse displacement from the axis, \ie

\be
B=B^{\prime}r,
\ee

\ni then, defining the \gb confinement time, $T_{\nabla B}$ to be the
time for an electron to drift half the length of a quad, $L_Q$, we get

\be
T_{\nabla B}={eB^{\prime}r^2L_Q\over 4 E c},
\ee

\ni where $E$ is the thermal energy of the electron. This
becomes

\be
T_{\nabla B}={\bar B^{\prime}\bar r^2\bar L_Q\over 4{\cal E}}\,\, \mu{\rm sec},
\ee

\ni where $\bar B^{\prime}$ is the field gradient in T/m, $\bar r$
is the radius in cm, ${\cal E}$ is the energy in eV,
 and $\bar L_Q$ is the quad length in cm. As an example typical
of the PSR,
if we take $\bar B^{\prime}=3.7$, $\bar r=2.5$, and $\bar L_Q=47$, then

\be
T_{\nabla B}={272\over {\cal E}}\,\, \mu{\rm sec}.
\ee

\ni Note that this is an overestimate of
the  drifts since the actual drift reverses sign as the electrons
move out along the magnetic field lines toward the poles. 
If the electrons only have energies that are a few hundred eV
then the confinement time is tens to hundreds of turns and is 
probably longer than the growth time for the
\ep instability.

The \exb drift velocity is given by

\be
V_{E\times\!B}={E_{\perp}\times B\over \vert B\vert ^2}c
\ee

\ni The electric field is due to the proton beam and to any electrons
that are present.
 The electric potential due to the
proton beam alone is given by

\be
e\phi ={2eI\over \beta c},
\ee

\ni where $\beta$ is the beam velocity scaled by the speed of light
and $I$ is the (time-dependent) beam current. The beam current
varies by $100\%$ in one revolution period (the beam is bunched), but
we will estimate drifts using the average current and resulting field.
Note that electrons
spend a lot of time near the magnetic mirror points where we expect
that the $E$ field will mostly be parallel to the $B$ field and
will give rise to only small drifts.
Nevertheless, the \exb  drift velocity
due to this term alone, assuming it acts all the time,
 would give an electron confinement time of

\be
T_{E\times B}=8.34{\beta \bar B^{\prime}\bar r^2\bar L_Q\over {\cal I}}\,\, 
{\rm nsec},
\ee

\ni where the current, ${\cal I}$, is measured in amps.
if we again take $\bar B^{\prime}=3.7$, $\bar r=2.5$, ${\cal I}=10$,
 and $\bar L_Q=47$, then $T_{E\times B}=761$ nsec, \ie electrons
would be confined for several turns, even with this overestimate
of the \exb drift. With a more realistic calculation, including
the full orbit dynamics of the electrons and the reverse
drifts that occur when only the electrons are present, it is likely that the
electrons will be confined for many turns.

Electrons to the left and right of the beam, horizontally, are free to move vertically
(initially)
until they move out radially along the field line to a region of greater field strength. 
Electrons above and below the beam, vertically, are free to move horizontally (initially)
until they move out radially along a field line
to a region of greater field strength. A complete model
of the electron motion is very complicated, but a simple model will suffice to treat
the motion of the center of mass of the electrons for oscillations near the beam
axis. Let us note that for electrons that can move vertically, \ie those to the left
and right of the beam, the restoring mirror force exactly vanishes at zero vertical position
and the restoring force reverses sign there.
For electrons that can move horizontally, \ie those above and below
 the beam, the restoring mirror force exactly vanishes at zero horizontal position
and the restoring force reverses sign there. Thus, in both transverse directions
we should expect
the restoring potential for an electron to be approximately a harmonic oscillator
potential near the axis. To see this in more detail, let us begin with the
equation from Krall and Trivelpiece, \cite{krall}, for the equation of motion along a field
line of a particle in a magnetic field

\be
{d^2 s\over dt^2}\approx -{v_{\perp 0}^2\over 2B_0}{\pl B\over \pl s},
\ee

\ni where $s$ is the distance along the field line, $v_{\perp 0}$ is the initial
value of the transverse velocity, and $B_0$ is the initial value of the magnitude 
of the magnetic field. The components of the quadrupole field are

\be
B_x=B_0^{\prime}y,
\ee

\be
B_y=B_0^{\prime}x.
\ee

\ni We thus see that

\be
{\pl B\over \pl s}={2B_0^{\prime}xy\over x^2+y^2}.
\ee

\ni From this we see that a particle that starts at $x=x_0$, $y=0$
satisfies the approximate equation

\be
{d^2y\over dt^2}\approx -({B_0^{\prime}v_{\perp 0}^2\over B_0x_0})y,
\ee

\ni \ie it is approximately a harmonic oscillator with a squared angular frequency
of

\be
\omega _m^2={B_0^{\prime}v_{\perp 0}^2\over B_0x_0}.
\ee

\ni But $B_0\approx B_0^{\prime} x_0$, so

\be
\omega _m^2\approx {v_{\perp 0}^2\over x_0^2}.
\label{gmax}
\ee

\ni It thus appears that
modeling the mirror trapping of the electrons by a harmonic oscillator potential, but
with a large spread in oscillation frequencies, should be
a fairly good approximation.

\section{Dipole Model of the \textit{e-p} Instability}

In this section, in order to find thresholds and growth rates, we will present a 
simple theory of the \emph{e-p} instability.
The model for the linear theory of the instability in this section 
that we use is similar to the theory of Keil and Zotter, \cite{keil-zotter}. 
We model the proton beam by the beam centroid at each azimuthal 
position around the ring.
The background electrons have a complex distribution both in physical
and in velocity space determined by their formation, capture in the
quadrupoles, interaction with the proton beam, and loss, as discussed in the previous section.
We cannot hope to accurately model all of these effects in an analytic theory; we will
simply assume that the electrons have a distribution in the squared magnetic
bounce frequency, $g_m = \omega _m^2 $, and that at each bounce frequency those electrons
are described by their centroid position, with electrons at a different bounce frequency
having a different centroid.
We assume the proton beam moves at a constant 
azimuthal velocity around the ring and is subject to a constant transverse focusing 
force that produces betatron oscillations at the betatron frequency, \ie we make the
smooth approximation, \cite{pjc2}. We only 
model proton beam and electron motion in one transverse direction. The protons and 
electrons are assumed to interact with each other via a force that is linear in the 
relative displacement of the centroids of the protons and electrons. 
The equations of motion for the centroids are thus 
given by

\begin{equation}
(\frac{\partial }{\partial t}+\omega _{0}\frac{\partial }{\partial 
\theta })^{2}y_{p}+\Gamma _d (\frac{\partial }{\partial t}+\omega _{0}\frac{\partial }{\partial 
\theta })y_{p}=-\omega ^{2}_{\beta }y_{p}+\omega 
_{p}^{2}(Y_{e}-y_{p})
\end{equation}
\label{sec: proton-eq}

\begin{equation}
\frac{\partial ^{2}y_{em}}{\partial t^{2}}+ \omega ^{2}_{m}y_{em}=\omega ^{2}_{e}(y_{p}-y_{em})
\end{equation}

\ni where \( y_{p}(\theta ,t) \) is the proton centroid position at an 
azimuth, \( \theta  \), around the machine and time, \( t \),  \( \omega _{0} \) is the proton 
beam angular revolution frequency in the machine, and \( \omega _{\beta } \) is the 
angular betatron frequency of the protons. The proton beam centroid only
responds to the net electron centroid position, $Y_{e}$, which is given by

\be
Y_{e}=\int f(g_m)y_{em}(\theta ,t)dg_m,
\label{net-center}
\ee

\ni where $f(g_m)$ is the equilibrium distribution function of electrons in the
squared bounce frequency and $y_{em}(\theta ,t)$ is the centroid of electrons
with a particular bounce frequency.
The coupling frequencies \( \omega _{p} 
\) and \( \omega _{e} \) are given by

\begin{equation}
\omega ^{2}_{e}=\frac{2N_{p}r_{e}c^{2}}{\pi b(a+b)R}
\end{equation}
 
\begin{equation}
\omega ^{2}_{p}=(\frac{F m_{e}}{\gamma m_{p}})\omega ^{2}_{e}
\end{equation}
with \( N_{p} \) the number of protons in the machine, \( r_{e} \) 
the classical electron radius, \( c \) the velocity of light, \( \gamma  \) the 
relativistic factor of the proton beam, \( a \) and \( b \) the sizes of the 
proton beam, \( F \) the neutralization fraction of electrons, and \( R \) the 
effective radius of the ring. Note that the inter-species force is assumed to 
depend linearly on the distance between the beam centroids; this is approximately 
correct for small amplitudes of oscillation, but clearly fails at larger 
oscillation amplitudes.

Also note that we have inserted a linear damping term with coefficient
$\Gamma _d $ into the proton equation to account
for the chromatic spread  in proton revolution frequencies;
the different revolution frequencies will give different longitudinal
velocities which will Landau damp the transverse oscillations.
A more extensive model would have the proton beam described by
a distribution function in the azimuthal direction and take into
account the Landau damping due to the spread in azimuthal velocities.
The approximation we have adopted mimics this damping and
has the same functional dependence
as the result of this more extensive model (see below), \ie
 the damping depends on 1) the energy spread, 2) the momentum compaction
factor, and 3) the mode number (through the derivative in
the damping term). 
 Thus, this damping term will give rise to the correct qualitative
behavior with the correct functional dependencies,
 \ie damping of off-axis oscillations as they phase-mix away. 
We can estimate
this damping rate of transverse oscillations
due to this spread
to be the chromatic fractional tune spread times the betatron frequency.
Note that the chromatic fractional tune spread is just the chromaticity times
the energy spread, i.e. it measures the longitudinal velocity spread and
its influence on the transverse oscillations. 
We do not include the transverse tune spread due to space charge and machine 
nonlinearities because we are using a dipole model and the centroid motion
of the protons does not depend on these terms.

\be
\Gamma _d \sim ({\Delta \nu \over \nu}){\omega _{\beta}\over 2\pi}.
\label{damp-exp}
\ee

\ni Because we are using an unbunched beam model, i.e. the smooth 
approximation, the average neutralization around the ring will
be smaller than the neutralization in the quadrupoles by roughly
the ratio of the ratio of total quadrupole length to the
ring circumference; thus the neutralization fraction in a
quadrupole will be about $20$ times $F$ since quadrupoles are about
$10$\% of the circumference and only about half the electrons can
move vertically. 

We have seen in the context of the drift
space instability model, \cite{pjc}, that bunching doesn't have a large
effect on the instability, and we assume the same to be true here.
There seems to be no simple way to incorporate bunching; a moderately realistic 
model would result in a dispersion equation which would be an infinite
matrix eqation with all unbunched beam modes coupled. The unbunched
beam model of this paper would then be just the diagonal approximation
to this matrix equation. It is likely that an extensive numerical
investigation would be required to resolve the behavior.

The above model is overly simplified, but contains most of 
 the important physics. It will break down, of course, if the
electron loss rate is too high. Of couse, we are also
assuming that the background electron density, on average, is constant so that
if electron generation and loss rates fluctuate rapidly our model should fail. 

The model of Bosch, \cite{bosch}, for the effect of beam gaps on the
trapped ion instability in an electron ring also considers the effect
of a large spread of (ion) oscillation frequencies on the instability,
and his formulation is similar to ours.

If we 
assume that the perturbations have a dependence on time and angle proportional to 
\( e^{-i\omega t+in\theta } \), then the equations become

\be 
(-(\omega -n\omega _{0})^{2}-i\Gamma _d (\omega -n\omega _{0})+\omega ^{2}_{\beta }+\omega 
^{2}_{p})y_{p}=\omega ^{2}_{p}Y_{e}
\label{proton-eq2}
\ee

\be
(\omega ^{2}_{e}+\omega ^{2}_{m}-\omega ^{2})y_{e}=\omega ^{2}_{e}y_{p}.
\label{electron-eq2}\ee

\ni Solving equation \ref{electron-eq2} for $y_e$ and using equations \ref{net-center}
and \ref{proton-eq2} we find

\be
((\omega -n\omega _{0})^{2}+i\Gamma _d (\omega -n\omega _{0})-\omega ^{2}_{\beta }-\omega 
^{2}_{p})=-\omega ^{2}_{e} \omega ^{2}_{p}\int {f(g_m)\over g_m + \omega ^{2}_{e} - \omega ^{2}}dg_m,
\label{dispersion1}
\ee

\ni where we have used the definition of $g_m=\omega^2_m$. We have to deal with the singularity in the
integral on the right hand side of this equation. We adopt the Landau prescription, see \cite{krall},
where the integral is replaced by the principal value plus $\pi i$ times the residue at the pole;

\be
\int {f(g_m)\over g_m + \omega ^{2}_{e} - \omega ^{2}}dg_m= \dashint {f(g_m)\over g_m + \omega ^{2}_{e} - \omega ^{2}}dg_m
+\pi i f(\omega ^2 - \omega ^2_e). 
\ee

\ni Equation \ref{dispersion1} thus becomes

\begin{eqnarray}
((\omega -n\omega _{0})^{2}+i\Gamma _d (\omega -n\omega _{0})-\omega ^{2}_{\beta }-\omega 
^{2}_{p})& = & -\omega ^{2}_{e} \omega ^{2}_{p}\dashint {f(g_m)\over g_m + \omega ^{2}_{e} - \omega ^{2}}dg_m
\nonumber \\
 & & -\pi i \omega ^{2}_{e} \omega ^{2}_{p}f(\omega ^2 - \omega ^2_e),
\label{dispersion2}
\end{eqnarray}

\ni where the bar through the integral sign indicates principal value.
This is the dispersion relation for the two-stream mode. To solve it we have to
specify the distribution function of electron bounce frequencies, $f$. Of course, 
there should be no electrons in the `loss-cone', \ie at zero $\omega ^2_m$, but 
otherwise the detailed distribution depends on  
 their formation, capture in the
quadrupoles, interaction with the proton beam, and loss. We will simply take 
one distribution as an example, one in which the distribution is constant
between a minimum squared bounce frequency and a maximum squared bounce frequency;
\ie

\begin{eqnarray}
f(g_m)&=& {1\over g_{max} -g_{min}}\; \; \; g_{min}\leq g_m \leq g_{max}, \nonumber \\
 & & 0 \;\;\;\;\;\;\;\;\;\;\;\;\;\;\;\;\;\;\;\;\;\;\;\;\;\;\;\;\;\; {\rm otherwise}
\end{eqnarray}

\ni With this distribution the dispersion equation, \ref{dispersion2}, becomes

\begin{eqnarray}
(\omega -n\omega _{0})^{2}& = &\omega ^{2}_{\beta }+\omega 
^{2}_{p}-i\Gamma _d (\omega -n\omega _{0}) \nonumber \\
  & &  -{\omega ^{2}_{e} \omega ^{2}_{p}\over g_{max} -g_{min}}
(\ln ({g_{max}-\omega ^2 +\omega ^2_e \over \omega ^2 - \omega ^2_e -g_{min}})+ \pi i) 
\label{dispersion3}
\end{eqnarray}

\ni Though this is a transcendental equation and can't be solved exactly, we note that
the coefficient of the logarithmic term is small and the logarithm varies slowly,
so we can simply solve iteratively. The remainder of the equation is a quadratic for
$\omega -n\omega _{0} $ and the resulting approximate solution is

\begin{eqnarray}
\omega & \approx & n\omega _0 -{i\Gamma _d \over 2} \nonumber \\
 & & \pm {1\over 2}\bigl[ 4(\omega ^2_{\beta} + \omega ^2_p)-\Gamma _d^2\nonumber \\
 & &  -{4 \omega ^{2}_{e} \omega ^{2}_{p}\over g_{max} -g_{min}}
(\ln ({g_{max}-(n\omega _0)^2 +\omega ^2_e \over (n\omega _0)^2 - \omega ^2_e -g_{min}})+ \pi i) 
\bigr]^{1\over 2} 
\end{eqnarray}

We note the damping due to the phase mixing term, as expected, and the usual
upper and lower sidebands. Note that we have taken
$\omega  \approx  n\omega _0$ inside the logarithm because the
mode numbers are usually rather high $30 - 50$ and this is a good (few percent)
approximation for the real part of the frequency. 
Let us expand just the imaginary term under the square root
to find the damping and growth rates. For convenience define the real frequency shift to
be

\be
\omega ^2_s \equiv  {1\over 4}\bigl[ 4(\omega ^2_{\beta} + \omega ^2_p)-\Gamma _d^2\nonumber \\
  -{4 \omega ^{2}_{e} \omega ^{2}_{p}\over g_{max} -g_{min}}
(\ln ({g_{max}-(n\omega _0)^2 +\omega ^2_e \over (n\omega _0)^2 - \omega ^2_e -g_{min}})) \bigr]
\ee

\ni Note that to a good approximation $\omega _s \approx \omega _{\beta}$. Expanding the imaginary term in the square root we get

\be
\omega  \approx  n\omega _0 -{i\Gamma _d \over 2} 
\pm \omega _s(1   - {\pi i \omega ^{2}_{e} \omega ^{2}_{p}\over 2 \omega ^2_s (g_{max} -g_{min}))}
 ) 
\ee

\ni Note that the upper side band (plus sign) is always damped, but that the lower
side band can be unstable if

\be
{\pi \omega ^{2}_{e} \omega ^{2}_{p}\over  \omega _s (g_{max} -g_{min}))} > \Gamma_d,
\label{thresh1}
\ee

\ni with growth rate given by

\be
\gamma _{\rm growth}={\pi \omega ^{2}_{e} \omega ^{2}_{p}\over 2 \omega _s (g_{max} -g_{min}))} - {\Gamma_d \over 2}.
\label{grate1}
\ee

A number of modes in lower side bands can be unstable,
limited only by the condition
$\omega_e^2+g_{min}<\omega^2<\omega_e^2+g_{max}$, with roughly equal
growth rates (there is some weak dependence on mode number in $\omega _s$)
and this is consistent with experiments where multiple modes are usually
observered, \cite{macek2}.

\subsection{Example}

 Let us look at 
an example typical of the PSR; let us take

\[
a=b=1.8\; {\rm cm},\]
\[
\omega _{0}=2\pi *2.8\; {\rm MHz},\]
\[
\omega _{\beta }=2.2*\omega _{0},\]
\[
2\pi R=89\; {\rm m}\]

\ni If we express the number of particles in the ring as

\be
N_p={\cal N}\times 10^{13},
\ee

\ni then we can compute

\be
\omega ^2_e=1.76 * {\cal N}* 10^{17}\; {\rm  sec}^{-2},
\label{oe2}
\ee

\ni and

\be
\omega ^2_p=0.5182*  F * {\cal N}* 10^{14}\; {\rm  sec}^{-2}.
\label{op2}
\ee

\ni 
In the PSR the measured vertical chromaticity is about $-1.68$ and the
energy spread (typical conditions) is about $0.5\%$ so we take the
chromatic tune spread to be about $0.009$, \ie a fractional
tune spread of $0.43$\%, then

\be
\Gamma _d \approx 0.0252 * 10^{6}\; {\rm  sec}^{-1}.
\label{gd}
\ee

\ni We take the frequency shift to be

\be
\omega _s \approx \omega_{\beta} = 3.9 * 10^7 \; {\rm sec}^{-1}.
\label{shift}
\ee

\ni To estimate $g_{max}$ we use equation \ref{gmax}, setting the maximum transverse energy
to the beam potential; the result is

\be
g_{max}\approx 3.48*  {\cal N}*  10^{17} \; {\rm sec}^{-2},
\label{gmval}
\ee

\ni where we used $x_0 \approx a = 1.8$ cm. Note that we simply ignore $g_{min}$, \ie assume it is zero; it only
modifies our results by a small factor.

If we evaluate the threshold condition, equation \ref{thresh1}, using equations \ref{oe2}, \ref{op2}, \ref{gd},
\ref{shift}, and \ref{gmval}  we find the criterion for instability to be

\be
F*{\cal N} > 0.188;
\ee

\ni in other words, once the product of the particle number (times $10^{13}$) and percent
neutralization is about $19.0$, we can expect instability. Recall that the neutralization
fraction in quadrupoles will be about $20$ times higher than $F$ since quadrupoles are only
about $10$\% of the ring and only half the electrons can move vertically. 
At threshold the growth time is infinite, but if, for 
simplicity, we assume that we are a factor of $2$ above the threshold, $F*{\cal N} = 9.4* 10^{-2}$, then using
\ref{oe2}, \ref{op2}, \ref{gd}, \ref{shift}, and \ref{gmval} in equation
\ref{grate1} we find

\be
\gamma _{\rm growth} \approx 12.6 \; {\rm KHz},
\ee

\ni \ie a growth time of about $222$ turns. 
These estimates are only intended to show that the results seem to be within a factor of
two or three of the observations and that the theory is thus a possible explanation of 
the observed instability.

\section{Discussion}

Our results show that electrons trapped in quadrupoles are a plausible explanation of
the two-stream instability observed in the PSR. The growth times found are considerably
closer to the observed values than the linear growth times derived from the instability treatment
for electrons in drift spaces, \cite{pjc}. The reason for this is that the electrons confined in 
quadrupoles have a very large frequency spread due to the wide variation in magnetic bounce
frequencies as compared to those in drift spaces which have only a very small spread in
space charge confinement bounce frequencies. Thus, many fewer electrons are resonant at a particular
frequency. 

In addition, if the instability is due to electrons trapped in quadrupoles, then the transverse
momentum kick given to the protons is easily explained; the momentum is transferred from
the quadrupoles via the electrons, rather than having to be transferred only from electrons,
as in the drift space theory.

Clearly a great deal more work can be done to refine this model. A kinetic description
of the proton beam could be used, and would give a more sensitive dependence of
the phase-mixing damping that depends on the detailed proton distribution.
An investigation of different electron distribution functions, perhaps motivated
by detailed simulation of electron formation and capture dynamics, would give
threshold and growth rate estimates that are better founded than those in this
note. The formulation of a bunched beam model would be considerably more
difficult, but might be worthwhile. Finally, a composite model with both
drift space electrons and quadrupole trapped electrons would be very
difficult to analyze but might be necessary to fit all observations in real
machines.

\end{document}